\magnification=1200 \baselineskip=24 truept plus1pt minus1pt
\parskip=9 true pt
\hoffset=0.5 true in \hsize=6.2 true in
\nopagenumbers \pageno =1 \headline{\hss\tenrm--\ \folio\ --\hss}
\def\section#1{\goodbreak\medbreak\noindent{\tenbf #1}\nobreak\smallskip}
\def\subsection#1{\goodbreak\smallbreak\noindent{\bf#1}\nobreak\smallskip}
\def\ref#1#2#3#4#5#6{\medbreak\noindent\item{#1}{#2},\ {#3}\
{\it #4}{#5}{#6}}

 \font\bigbold=cmbx10 scaled\magstep1

\def\romone{\uppercase\expandafter{\romannumeral 1}}
\def\romtwo{\uppercase\expandafter{\romannumeral 2}}
\def\romtre{\uppercase\expandafter{\romannumeral 3}}
\def\romfor{\uppercase\expandafter{\romannumeral 4}}
\def\romfiv{\uppercase\expandafter{\romannumeral 5}}
\def\romsix{\uppercase\expandafter{\romannumeral 6}}
\def\romsev{\uppercase\expandafter{\romannumeral 7}}
\def\romeig{\uppercase\expandafter{\romannumeral 8}}
\def\romnin{\uppercase\expandafter{\romannumeral 9}}

\def\caption #1! #2!{\noindent{\bf #1\enspace}{\sl #2}}

\def\toclin1#1#2{\line{\noindent #1\ \dotfill\ #2}}
\def\toclin2#1#2{\line{\indent #1\ \dotfill\ #2}}


\input epsf

\centerline{\bigbold Fractional Parallel Plate DB Waveguides}
\centerline{\bigbold Using Fractional Curl Operator}

\vskip 0.5 in
 \centerline{Akhtar Hussain$^1$, Sajid Abbas Naqvi$^2$, $^*$Ahsan Illahi$^3$, and Q. A. Naqvi$^4$}
\vskip 0.3 in

\centerline{Electronics Department,  Quaid-i-Azam University
Islamabad} \centerline{$^*$ Physics Department,  Allama Iqbal Open
University Islamabad}

\vskip 0.3 in
 \centerline{{\bf E-mail}: $^1$akhtar\_h@yahoo.com,
$^2$nsajidabbas@yahoo.com,}  \centerline{$^3$ahsanilahi@gmail.com
$^4$qaisar@qau.edu.pk}

\vskip 1.2 in
  \noindent \centerline{\bf Abstract}\vskip 0.5 cm
 \noindent{ DB boundary conditions have been simulated in terms of perfect electric conductor (PEC) boundary
 for transverse electric modes and perfect magnetic conductor (PMC) for transverse magnetic modes.
 Electric and magnetic fields inside the dielectric region of a parallel plate DB boundary
 waveguide have been derived and fractional curl operator has been utilized to study
 the fractional parallel plate DB waveguide. Field behavior as well as transverse impedance of the guide walls have been
 studied with respect to the fractional parameter describing the
 order of fractional curl operator. The results are also compared
 with the corresponding results for fractional waveguides with PEC walls.

\vfill\eject

\section{1. Introduction }

Fractional calculus is a branch of mathematical analysis that
studies the possibility of taking real number powers or complex
number powers of the differentiation and integration operators~[1].
It fills in the gaps of classical calculus, so in this respect
traditional calculus may be taken as a special case of the
fractional calculus. Using fractional calculus, scientists and
engineers have been interested in exploring the potential utilities
and possible physical implications of mathematical machinery of the
subject of fractional calculus, that is, fractional derivatives and
fractional integrals~[2]. Fractional derivatives and integrals are
mathematical operators involving differentiation and integration of
arbitrary (non-integer) real or complex orders such that
$d^n{f(x)}/d{x}^n$ and $d^{-m}f(x)/dx^{-m}$ respectively, where $m$
and $n$ can be taken as non-integer real or even complex number. It
has been  demonstrated that these mathematical operators are
interesting and useful tools in various disciplines of science and
engineering~[3-5].

 Electromagnetic theory has a driving role in
the modern world of science and engineering.  Maxwell equations
encapsulate all this revolutionary discipline, whose solutions have
a great importance in current research and development activities.
Fifteen years ago, Engheta particularly focused on finding out what
possible applications and/or physical role, the mathematical
operators of fractional calculus can have in electromagnetic
theory~[6-11]. He applied the concept of fractional
derivatives/integrals to certain electromagnetic problems, and
obtained interesting results and ideas showing that these
mathematical operators are interesting and useful mathematical tools
in electromagnetic theory. Some of these ideas include the
mathematical link between the electrostatic image methods for the
conducting sphere and the dielectric sphere~[6], fractional
solutions for the scalar Helmholtz equation~[7,10], electrostatic
fractional image methods for perfectly conducting wedges and
cones~[8], and the novel concept of fractional multipoles in
electromagnetism~[9].

Other researchers belonging to the area of electromagnetics have
also contributed to this topic. Tarasov proved that the
electromagnetic fields in dielectric media, whose susceptibility
follows a fractional power-law dependence in a wide frequency range,
can be described by differential equations with time derivatives of
noninteger order~[12,13]. Fractional dimensional space represents an
effective physical description of confinement in low-dimensional
systems. The concept of fractional calculus to obtain the solution
of electrostatic problem in fractional dimensional space, for the
fractional order multipoles, is utilized by Muslih and Baleanu~[14].
They also introduced the form of fractional scalar potential by
using the solutions of Laplace's equation in fractional dimensional
space~[15]. They derived  potential of charge distribution in
fractional space  using Gegenbauer polynomials. According to Zubair
et al.,  solutions of Helmholtz equation in fractional space can
describe the complex phenomenon of wave propagation in fractal
media. With this view, they established a generalized Helmholtz
equation for wave propagation in fractional space and found its
analytical solution~[16-21].

Fractionalization of ordinary derivative and integral operators
motivated the researchers to  find out the possible
fractionalization of  other operators and their use in
electromagnetics. Engheta fractionalized the kernel of an integral
transform and studied the paradigm of intermediate zones in
electromagnetism~[22,23]. Fractionalization of the curl operator, a
well known operator in vector calculus, was also introduced by
him~[24]. He used the fractional curl operator to find the new
solutions, which may be regarded as intermediate step between the
two given solutions, to the Maxwell equations.

For the sake of completeness, it is decided  first to reproduce the
meanings of the following concepts given by Engheta~[24]: what is
meant by fractionalization of a linear operator? How to
fractionalize a linear operator? That is, what is the mathematical
recipe to fractionalize a linear operator? Fractionalization of the
curl operator using this recipe and utilization of the fractional
curl operator to find intermediate/fractional solutions to the
Maxwell equations for ordinary medium has been discussed~[24] and
just results of their work are presented here. Time harmonic
dependency $\exp(-i\omega t)$ has been considered throughout the
paper.
\section{1.1. Fractional linear operator in electromagnetics}
\section{1.1.1. Conditions required to verify fractionalization }
Mathematical fractionalization of any problem requires two canonical
solutions of the problem under consideration and an operator that
can transform one canonical solution into the other.
Fractionalization of the connecting operator can reveal intermediate
solutions between the two canonical solutions. The conditions and
recipe for fractionalization  of a linear operator ${\underline
{\underline {\bf L}}}$ are reproduced below~[24]. It must be
mentioned that this recipe  has also been used to fractionalize the
Fourier transform~[25]. The new fractionalized operator ${\underline
{\underline {\bf L}}}^\alpha$ with fractional parameter $\alpha$,
under certain conditions, can be used to obtain the intermediate
cases between the canonical case 1 and case 2.

 A linear operator
${\underline {\underline {\bf L}}}$
 may be a fractional operator (i.e.,
${\underline {\underline {\bf L}}}^\alpha$) that provides  the
intermediate solutions to the original problems, if it satisfies the
following properties~[11,24].

 {\bf I.} For $\alpha =1 $, the fractional
operator ${\underline {\underline {\bf L}}}^\alpha$ should become
the original operator ${\underline {\underline {\bf L}}}$, which
provides us with case 2 when it is applied to case 1.

{\bf II.} For $\alpha =0 $, the operator ${\underline {\underline
{\bf L}}}^\alpha$ should become the identity operator ${\underline
{\underline {\bf I}}}$ and thus the case 1 can be mapped into
itself.

{\bf III.} For any two values $\alpha_1$ and $\alpha_2$ of
fractional parameter, ${\underline {\underline {\bf L}}}^\alpha$
have the additive property in $\alpha$, i.e.,
$${\underline {\underline {\bf L}}}^{\alpha_1}
. {\underline {\underline {\bf L}}}^{\alpha_2} = {\underline
{\underline {\bf L}}}^{\alpha_2} . {\underline {\underline {\bf
L}}}^{\alpha_1} = {\underline {\underline {\bf
L}}}^{\alpha_1+\alpha_2}\eqno(1.1)
$$

{\bf IV.} The operator ${\underline {\underline {\bf L}}}^\alpha$
 should commute with the operator
involved in the  mathematical description of the original problem.
\section{1.1.2. Recipe for fractionalization }
It is assumed that,  present discussion is about a class of linear
operators (or mappings) where the domain and range of any linear
operator of this class are similar to each other and have the same
dimensions.
That is, $ {\underline {\underline {\bf L}}}^\alpha : C^n
\rightarrow C^n$ where ${C}^n$ is a $n$ dimensional vector space
over the field of complex numbers. Once a linear operator such as
${\underline {\underline {\bf L}}}$ is given, the recipe for
constructing the fractional operator ${\underline {\underline {\bf
L}}}^\alpha$ can be described as follows~[24].

\item{\bf 1.} One finds the eigenvectors and eigenvalues of the
operator ${\underline {\underline {\bf L}}}$ in the space $C^n$ so
that $ {\underline {\underline {\bf L}}} .{\bf A}_m = a_m {\bf A}_m$
where ${\bf A}_m$ and $a_m$ for $m=1,2,3....n$, are the eigenvectors
and eigenvalues of the operator ${\underline {\underline {\bf L}}}$
 in space $C^n$ respectively.

\item{\bf 2.} Provided ${\bf A}_m$s form a complete orthogonal basis in the
space $C^n$, any vector in this space can be expressed in terms of
linear combination of ${\bf A}_m$. Thus an arbitrary vector ${\bf
G}$ in space $C^n$ can be written as
$${{\bf G }= {\sum}_{m=1}^n  g_m{\bf A}_m} \eqno(1.2)$$
where $g_m$s are co-efficients of expansion of ${\bf G}$ in terms of
${\bf A}_m$s.

\item{\bf 3.} Having obtained the eigenvectors and eigenvalues of the
operator ${\underline {\underline {\bf L}}}$, the fractional
operator ${\underline {\underline {\bf L}}}^\alpha$ can be seen to
have the same eigenvectors ${\bf A}_m$s but with the eigenvalues as
$(a_m)^\alpha$, i.e.,
$$ {\underline {\underline {\bf L}}}^\alpha.{\bf A}_m = (a)_m^\alpha {\bf A}_m \eqno(1.3)$$
When this fractional operator ${\underline {\underline {\bf
L}}}^\alpha$
 operates on an arbitrary vector ${\bf G}$ in the space $C^n$, one gets
$$\eqalignno{{\underline {\underline {\bf L}}}^\alpha
.{\bf G }&= {\underline {\underline {\bf L}}}^\alpha {\sum}_{m=1}^n
g_m{\bf A}_m\cr &= {\sum}_{m=1}^n g_m {\underline {\underline {\bf
L}}}^\alpha .{\bf A}_m= {\sum}_{m=1}^n g_m (a_m)^\alpha {\bf A}_m
&(1.4)}$$ The above equation essentially defines the fractional
operator ${\underline {\underline {\bf L}}}^\alpha$ from the
knowledge of operator ${\underline {\underline {\bf L}}}$ and its
eigenvectors and eigenvalues. In the next section, above recipe has
been applied to fractionalize the curl operator.
\section{ 1.1.3. Fractional curl operator and Maxwell equations}
Consider a three-dimensional vector field ${\bf F}$ as a function of
three cartesian space coordinates~$(x,y,z)$. Curl of this vector can
be written as
$$\eqalignno{ {\rm curl} {\bf F} &= \left({\partial{F}_z \over \partial y}
-{\partial{F}_y \over \partial z}\right){\hat {\bf x}}+
\left({\partial{F}_x \over \partial z}-{\partial{F}_z\over
\partial x}\right){\hat {\bf y}}+\left({\partial{F}_y \over \partial x}-{\partial{F}_x \over \partial y}\right){\hat
{\bf z}}&(1.5)}$$ where $F_x$, $F_y$, $F_z$ are the cartesian
components of vector ${\bf F}$ and ${\hat {\bf x}}$, ${\hat {\bf
y}}$, ${\hat {\bf z}}$ are the unit vectors in the space domain.
Assuming that spatial Fourier transforms of both the vector
functions (${\bf F}$ and ${\rm  curl{\bf F}}$) exist, the Fourier
transform of these two vectors can be written as
$$\eqalignno{{\cal {F}}_k \left\{ {\bf F}(x,y,z)\right\}&= {\tilde {\bf F}}(k_x,k_y,k_z)\cr
&=\int_{-\infty}^{\infty}\int_{-\infty}^{\infty}\int_{-\infty}^{\infty}
{\bf F}(x,y,z)\exp{(-ik_xx-ik_yy-ik_zz)} dx\,dy\,dz \,\,\, &(1.6)\cr
{\cal { F}}_k \left\{ {\rm curl} {\bf F}(x,y,z)\right\}
&=\int_{-\infty}^{\infty}\int_{-\infty}^{\infty}\int_{-\infty}^{\infty}
{\rm curl}{\bf F}(x,y,z)\exp{(-ik_xx-ik_yy-ik_zz)} dx\,dy\,dz\cr
 &=i{\bf k}\times{\tilde {\bf F}}{(k_x,k_y,k_z)}&(1.7)}$$
where a tilde over the vector ${\tilde {\bf F}}$ denotes the Fourier
transform of vector ${\bf F}$. Hence in the k-domain
$(k_x,k_y,k_z)$, the curl operator can be written as a cross product
of vector $i{\bf k}$ with the vector ${\tilde {\bf F}}$. In order to
fractionalize the curl operator, the cross product operator $( i{\bf
k} \times)$ in the k-domain serves as prerequisite. Thus
fractionalization of curl operator is equivalent to the
fractionalization of this cross product operator. With the recipe
described in the previous section, fractionalization of the cross
product operator as $(i{\bf k} {\times})^\alpha$ can be obtained in
the k-domain.

Engheta utilized the fractional curl operator to fractionalize the
principle of duality in electromagnetics~[24]. Fractionalization of
the principle of duality yields new solutions to the Maxwell
equations, which may be regarded as intermediate step between the
original solution and dual to the original solution and has been
termed as fractional dual solutions. In an isotropic, homogeneous,
and source free medium described by wavenumber $k$ and impedance
$\eta$, new set of solutions to the source-free Maxwell equations
may be obtained using the following relations~[24]
$$\eqalignno{{\tilde{ \bf E}}_{\rm fd} &= \left[{1\over (ik)^{\alpha}}(i{\bf
k}\times)^{\alpha}{\tilde{\bf E}}\right] &(1.8a)\cr \eta{\tilde{\bf
H}}_{\rm fd} &= \left[{1\over (ik)^{\alpha}}(i{\bf
k}\times)^{\alpha}\eta{\tilde{\bf H}}\right] &(1.8b)}$$ where ${\rm
fd}$ stands for the fractional dual.  Inverse Fourier transforming
these back into the $(x,y,z)$-domain, the new set of solutions are
obtained as
$$\eqalignno{{\bf E}_{\rm fd} &= \left[{1\over
(ik)^{\alpha}}{\rm curl }^{\alpha}{\bf E}\right] &(1.9a)\cr {\eta\bf
H}_{\rm fd} &= \left[{1\over (ik)^{\alpha}}{\rm curl
}^{\alpha}({\eta\bf H})\right] &(1.9b)}$$ From Eqs. (1.9), it can be
seen that for ${\alpha=0}$, $({\bf E}_{\rm fd},\eta{\bf H}_{\rm
fd})$ gives the original solutions whereas $({\bf E}_{\rm
fd},\eta{\bf H}_{\rm fd})$ gives dual to the original solution to
the Maxwell equations for ${\alpha=1}$. Therefore for all values of
${\alpha}$ between zero and unity, $({\bf E}_{\rm fd},\eta{\bf
H}_{\rm fd})$ provides the new set of solutions which can
effectively be regarded as intermediate solutions between the the
original solution and dual to original solution. These solutions are
also called the fractional dual fields as expressed with the
subscript ${\rm fd}$.
\section{1.4. Previous contributions}
As it has already been stated that concept of the fractional curl
operator and its utilization in electromagnetics was given by
Engheta~[24]. Naqvi and Rizvi  extended Engheta's work on fractional
curl operator by  determining  sources corresponding the fractional
dual solutions to the Maxwell equations. Results of their valuable
work show that surface impedance of the planar reflector, which is
intermediate step between the PEC and PMC, is anisotropic in
nature~[26]. Naqvi et al., further extended work on this topic by
finding fractional dual solutions to the Maxwell equations for
reciprocal, homogenous, and lossless chiral medium~[27]. Lakhtakia
pointed out that any fractional operator that commutes with curl
operator may yield fractional solutions~[23]. Naqvi and Abbas
studied the role of complex and higher order fractional curl
operators in electromagnetic wave propagation~[28]. They also
studied the fractional dual solutions in double negative (DNG)
medium~[29]. Veliev further extended  the work on the fractional
curl operator by finding the reflection coefficients and surface
impedance corresponding to fractional dual planar surfaces with
planar impedance surface as original problem~[30]. The work on this
topic entered into new era when concept of fractional transmission
lines, fractional waveguides, and fractional resonator in
electromagnetics were introduced~[31-40] and nature of the modes
supported by fractional dual waveguides and impedance of the walls
were addressed. Modelling of transmission of electromagnetic plane
wave through a chiral slab using fractional curl operator and
fractional dual solutions in bi-isotropic medium are also
available~[41,42]

After the introduction of nihility concept by Lakhtakia [43],
Tretyakov incorporated the nihility conditions to chiral medium and
proposed another metamaterial  termed as chiral nihility
metamaterial~[44,45]. Chiral nihility is a metamaterial with
following properties of constitutive parameters at certain
frequency~[45].
$$\epsilon\rightarrow 0,\qquad \mu\rightarrow 0,\qquad \kappa\neq 0 $$
Thus the resulting constitutive relations for isotropic chiral
nihility metamaterial reduces to
$$\eqalignno{{\bf D}&=i{\kappa} \sqrt{\epsilon_0\mu_0}{\bf H}&(1.15a)\cr
{\bf B}&= -i{\kappa} \sqrt{\epsilon_0\mu_0}{\bf E}&(1.15b)}$$

Tellegen nihility~[46] states that
$$\epsilon\rightarrow 0,\qquad \mu\rightarrow 0,\qquad \kappa\rightarrow 0,\qquad \chi \neq 0 $$
and corresponding expressions for constitutive relations for
Tellegen nihility metamaterial are
$$\eqalignno{{\bf D}&=\chi \sqrt{\epsilon_0\mu_0}{\bf H}&(1.16a)\cr
{\bf B}&= \chi  \sqrt{\epsilon_0\mu_0}{\bf E}&(1.16b)}$$ Study of
nihility/chiral nihility metamaterials is a topic of current
research by several researchers~[47-57]. Naqvi contributed many
research articles on chiral nihility and fractional dual solutions
in chiral nihility metamaterial~[51-57].

\section{1.3. DB boundary conditions}
Before the advent of idea 'DB boundary interface' proposed by
Lindell and Sihvola[58,59], all the known interfaces dealt with
tangential components of electric and magnetic fields. But the DB
boundary is analyzed on the basis of normal components of flux
densities ${\bf D}$ and ${\bf B}$ [ 59,60]. Waves polarized
transverse electric~(TE) and transverse magnetic~(TM) with respect
to the normal of the boundary are reflected as from perfect electric
conductor~(PEC) and perfect magnetic conductor~(PMC) planes~( i.e.,
DB interface behaves like PEC and PMC)~[61]. It is worth mentioning
here that, all of the previous conditions in electromagnetics are
associated to the electromagnetic field vectors ${\bf E}$ and ${\bf
H}$. The boundary conditions for DB interface may be written as
[59-65]
$$ \eqalignno{{\bf \hat n}.{\bf D}&=0&(1.17a)\cr
{\bf \hat n}.{\bf B} &=0&(1.17b)}$$ where ${\bf \hat n}$ is normal
vector to the interface.

Purpose of the current discussion is to extend the previous work on
fractional waveguides for waveguide with DB walls.
\section{2. General wave behavior along a guiding structure }
Consider a waveguide consisting of two parallel plates separated
by a dielectric medium with constitutive parameters $\epsilon$ and
$\mu$. One plate is located at $y=0$, while other plate is located
at $y=b$. The plates are assumed to be of infinite extent and the
direction of propagation is considered as positive z-axis.
Electric and magnetic fields in the source free dielectric region
must satisfy the following homogeneous vector Helmholtz equations
$$\eqalignno{\nabla^2 {\bf E}(x,y,z)+k^2{\bf E}(x,y,z)&=0 &(2.1a)\cr
\nabla^2 {\bf H}(x,y,z)+k^2{\bf H}(x,y,z)&=0 &(2.1b)}$$ where
$\nabla^2={\partial^2 \over \partial x^2}+{\partial^2 \over
\partial y^2}+{\partial^2 \over \partial z^2}$ is the Laplacian operator and $k=\omega\sqrt{\mu\epsilon}$ is the wave
number. Taking z-dependance as $\exp(i \beta z)$, equation~(2.1) can
be reduced to two dimensional vector Helmholtz equation as
$$\eqalignno{\nabla_{xy}^2 {\bf E}(x,y)+h^2{\bf E}(x,y)&=0 &(2.2a)\cr
\nabla_{xy}^2 {\bf H}(x,y)+h^2{\bf H}(x,y)&=0 &(2.2b)}$$ where
$h^2=k^2- \beta^2$, $\beta$ is the propagation constant.

 Since propagation is along z-direction and the waveguide dimensions are
considered infinite in xz-plane. So x-dependence can be ignored in
the above equations. Under this condition, equation~(2.2) becomes
ordinary second order differential equation as
$$\eqalignno{{d^2{\bf E}(y) \over dy^2}+h^2{\bf E}(y)&=0 &(2.3a) \cr {d^2{\bf H}(y) \over dy^2}+h^2{\bf H}(y)&=0 &(2.3b)}$$
As a general procedure to solve waveguide problems, the Helmholtz
equation is solved for the axial field components only. The
transverse field components may be obtained using the axial
components of the fields and Maxwell equations. So scalar
Helmholtz equations for the axial components can be written as
$$\eqalignno{{d^2{E_z} \over dy^2}+h^2{E_z}&=0 &(2.3c) \cr {d^2{H_z} \over dy^2}+h^2{H_z}&=0 &(2.3d)}$$
General solution of the above equations is $$\eqalignno{E_z&=a_n
\cos(hy)+b_n \sin(hy)&(2.3e)\cr H_z&=c_n \cos(hy)+d_n
\sin(hy)&(2.3f)}$$ where $a_n, b_n, c_n, {\rm ~and~} d_n$ are
constants and can be found from the boundary conditions.

 Using Maxwell curl equations, the transverse
 components  can be expressed in terms of longitudinal components $(E_z, H_z)$,
 that is
$$\eqalignno{E_x &={1\over h^2}\left(i \beta {\partial E_z\over \partial x}+ik{\partial  \eta H_z\over \partial y}\right)&(2.4a)\cr
E_y&={1\over h^2}\left(i \beta {\partial E_z\over
\partial y}-ik{\partial  \eta H_z\over \partial x}\right)&(2.4b)\cr
H_x &={1\over h^2}\left(i \beta {\partial  H_z\over
\partial x}-{ik\over \eta}{\partial  E_z\over \partial y}\right)&(2.4c)\cr
H_y&={1\over h^2}\left(i \beta {\partial H_z\over
\partial y}+{ik\over \eta}{\partial  E_z\over \partial x}\right)&(2.4d)}$$
where
$$\eqalignno{\eta=&\sqrt{\mu \over \epsilon} {\rm ~~~is~
impedence~ of~ the~ medium~ inside~ the~ guide}}$$ In the
proceeding part of this paper, parallel plate waveguide with DB
boundary walls has been considered and the fractional dual
solutions have been determined and analyzed.

\section{3. Parallel plate DB boundary walls waveguide}
A wave of general polarization propagating in positive z-direction
through a parallel plate waveguide can be written as a linear sum
of the transverse electric($TE^z$) modes and transverse magnetic
($TM^z$) modes. A DB boundary can be simulated as the boundary
which behaves like perfect electric conductor (PEC) boundary for
($TE^z$) modes and perfect magnetic conductor (PMC) boundary for
($TM^z$)modes. Therefore fields inside a parallel plate DB
waveguide may be obtained by linear superposition of two canonical
solutions which are transverse electric($TE^z$) mode solution for
PEC waveguide and transverse magnetic($TM^z$) mode solution for
PMC waveguide. Both the cases have been discussed separately.
\section{3.1 Fractional dual solutions of canonical cases}
\section{ \qquad  Case 1:   Transverse electric ($TE^z$) mode
propagation through a PEC waveguide} Let us first consider that
($TE^z$) mode is propagating through a PEC waveguide described in
section~2. For this mode, axial component of the electric field is
zero while for the magnetic fields it is given as in
equation~(2.3f). Using equations~(2.4), the corresponding
transverse components can be written as
$$\eqalignno{ E_x&=\left({ik \over h}\right)\left[-c_n \sin(hy)+d_n
\cos(hy)\right] &(3.1a) \cr H_y&=\left({i\beta \over
h}\right)\left[-c_n \sin(hy)+d_n \cos(hy)\right] &(3.1b)\cr E_y&=0
&(3.1c) \cr H_x &=0 &(3.1d)}$$ Using boundary conditions for PEC
boundary that is $E_{x,z}=0|_{y={0,b}}$ ,  we get the particular
solutions as
$$\eqalignno{ E_x&=\left({ik \over h}\right)\left[-C_n \sin(hy)\right] &(3.2a) \cr \eta H_y&=\left({i\beta \over
h}\right)\left[-C_n \sin(hy)\right] &(3.2b)\cr \eta H_z&=C_n
\cos(hy) &(3.2c)\cr E_y&=0 &(3.2d) \cr H_x &=0 &(3.2e) \cr {\rm
where} \qquad C_n &=c_n \eta \qquad\qquad h ={n\pi \over b
}\qquad\qquad {\rm n=1,2,3...}}$$ Re-introducing the z-dependance
$\exp(i \beta z)$ and writing equations~3.2 in exponential form we
can write electric and magnetic fields inside the dielectric
region as sum of two plan waves as
$$\eqalignno{ {\bf E}&={\bf E}_1+{\bf E}_2 &(3.3a) \cr
\eta {\bf H}&=\eta{\bf E}_1+\eta{\bf H}_2 &(3.3b) }$$ where $({\bf
E}_1, {\bf H}_1)$ are the electric and magnetic fields associated
with one plane wave and are given below
$$\eqalignno{{\bf E}_1 &= \left({C_n\over 2}\right) \left({k\over
h}{\hat {\bf x}}\right) \exp(-ihy+i\beta z)  &(3.4a)\cr \eta{\bf
H}_1 &= \left({C_n\over 2}\right)\left({\beta\over h}{\hat {\bf
y}}+{\hat {\bf z}}\right)\exp(-ihy+i\beta z) &(3.4b)}$$ while
$({\bf E}_2, \eta{\bf H}_2)$ are the electric and magnetic fields
associated with the second plane wave  and are given below
$$\eqalignno{{\bf E}_2 &= \left({C_n\over 2}\right) \left(-{k\over
h}{\hat {\bf x}}\right) \exp(ihy+i\beta z)  &(3.5a)\cr \eta{\bf
H}_2 &= \left({C_n\over 2}\right)\left(-{\beta\over h}{\hat {\bf
y}}+{\hat {\bf z}}\right)\exp(ihy+i\beta z) &(3.5b)}$$ \vfil\eject

This situation can be shown as in Figure~1. \vskip 2.0 in

\epsfxsize= 12 pt\hskip 0.3 in \epsfbox[5 5 25 45]{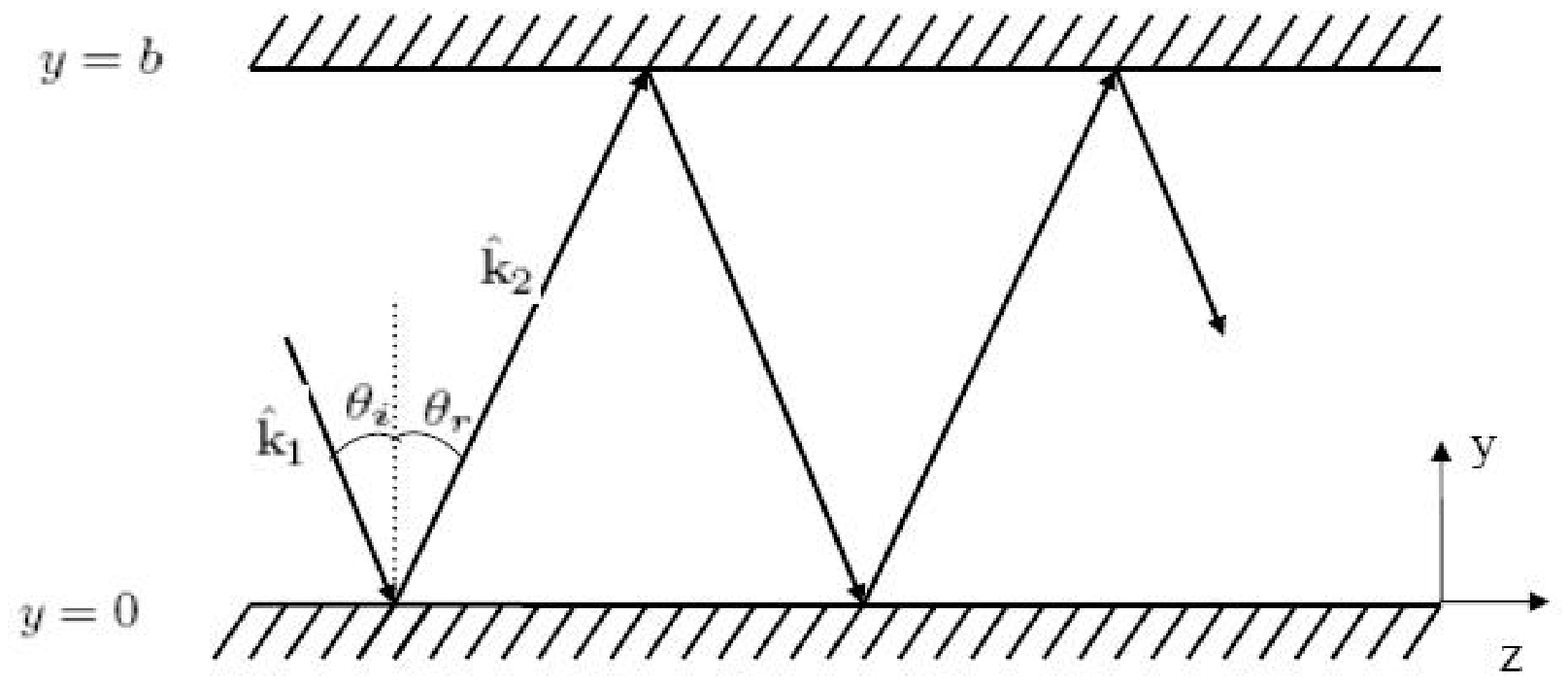}

\caption Figure 1 ! \rm Plane wave representation of the fields
inside the waveguide!

\hskip -0.3 in Once we have written electric and magnetic fields
inside the dielectric region in terms of two plan waves, recipe
for fictionalization [24,32] can be applied to get the fractional
dual solutions as
$$\eqalignno{{\bf E}_{\rm PECfd}^{\rm TE} &= C_n \left({k\over
h}\right)\left[  -i C_{\alpha}\sin \left(hy+{\alpha\pi\over
2}\right){\hat {\bf x}} +{\beta \over k} S_{\alpha}\cos
\left(hy+{\alpha\pi\over 2}\right) {\hat {\bf y}}\right. \cr
&\left.\qquad \qquad\qquad\qquad- i {h\over k} S_{\alpha}\sin
\left(hy+{\alpha\pi\over 2}\right) {\hat {\bf
z}}\right]\exp\left[i\left(\beta z+{\alpha\pi \over
2}\right)\right]\qquad\qquad&(3.6a)\cr \eta{\bf H}_{\rm
PECfd}^{\rm TE} &= C_n \left({k\over h}\right)\left[ -
S_{\alpha}\cos \left(hy+{\alpha\pi\over 2}\right){\hat {\bf x}}
-i{\beta \over k} C_{\alpha}\sin \left(hy+{\alpha\pi\over
2}\right) {\hat {\bf y}} \right. \cr &\left.\qquad
\qquad\qquad\qquad+ {h\over k} C_{\alpha}\cos
\left(hy+{\alpha\pi\over 2}\right) {\hat {\bf
z}}\right]\exp\left[i\left(\beta z+{\alpha\pi \over
2}\right)\right]\qquad\qquad&(3.6b)}$$ where
$$\eqalignno{C_{\alpha}&=\cos \left({\alpha\pi\over
2}\right)\cr S_{\alpha}&=\sin \left({\alpha\pi\over 2}\right) }$$

\section{ \qquad  Case 2:   Transverse magnetic ($TM^z$)
mode propagation through a PMC waveguide} Similar to the treatment
done in Case 1, using equation~(2.3e) and equations~(2.4), we can
write the results for transverse magnetic mode propagating through
a PMC waveguide as
$$\eqalignno{{\bf E}_{\rm PMCfd}^{\rm TM} &= A_n \left({k\over
h}\right)\left[  - S_{\alpha}\cos \left(hy+{\alpha\pi\over
2}\right){\hat {\bf x}} -i{\beta \over k} C_{\alpha}\sin
\left(hy+{\alpha\pi\over 2}\right) {\hat {\bf y}}\right. \cr
&\left.\qquad \qquad\qquad\qquad+ {h\over k} C_{\alpha}\cos
\left(hy+{\alpha\pi\over 2}\right) {\hat {\bf
z}}\right]\exp\left[i\left(\beta z+{\alpha\pi \over
2}\right)\right]\qquad\qquad&(3.6a)\cr \eta{\bf H}_{\rm
PMCfd}^{\rm TM} &= A_n \left({k\over h}\right)\left[ i
C_{\alpha}\sin \left(hy+{\alpha\pi\over 2}\right){\hat {\bf x}}
-{\beta \over k} S_{\alpha}\cos \left(hy+{\alpha\pi\over 2}\right)
{\hat {\bf y}} \right. \cr &\left.\qquad \qquad\qquad\qquad+ {i
h\over k} S_{\alpha}\sin \left(hy+{\alpha\pi\over 2}\right) {\hat
{\bf z}}\right]\exp\left[i\left(\beta z+{\alpha\pi \over
2}\right)\right]\qquad\qquad&(3.6b)}$$

\section{3.2 Fractional dual parallel plate DB waveguide}
Fractional dual solutions for the DB waveguide can be written by
taking linear sum of the fractional dual fields of the above two
cases as
$$\eqalignno{{\bf E}_{\rm fd}&={\bf E}_{\rm PECfd}^{\rm TE}+{\bf E}_{\rm PMCfd}^{\rm
TM}\cr \eta{\bf H}_{\rm fd}&=\eta{\bf H}_{\rm PECfd}^{\rm
TE}+\eta{\bf H}_{\rm PMCfd}^{\rm TM}}$$ which give
$$\eqalignno{{\bf E}_{\rm fd}=  \left({k\over
h}\right)\exp\left[i\left(\beta z+{\alpha\pi \over
2}\right)\right]& \bigl\{  -(A_n S_{\alpha}C_{y+\alpha}+iC_n
C_{\alpha}S_{y+\alpha}){\hat {\bf x}}\bigr. \cr  +&{\beta  \over
k} (C_n S_{\alpha}C_{y+\alpha}-i A_n C_{\alpha}S_{y+\alpha}) {\hat
{\bf y}} \cr+&\bigl. {h\over k} (A_n C_{\alpha}C_{y+\alpha}-iC_n
S_{\alpha}S_{y+\alpha}) {\hat {\bf z}}\bigr\}&(3.7a)}$$
$$\eqalignno{\eta{\bf H}_{\rm fd}=  \left({k\over h}\right)\exp\left[i\left(\beta
z+{\alpha\pi \over 2}\right)\right]& \bigl\{ -(C_n
S_{\alpha}C_{y+\alpha}-iA_n C_{\alpha}S_{y+\alpha}){\hat {\bf
x}}\bigr. \cr  -&{\beta  \over k} (A_n S_{\alpha}C_{y+\alpha}+iC_n
C_{\alpha}S_{y+\alpha}) {\hat {\bf y}} \cr+&\bigl. {h\over k} (C_n
C_{\alpha}C_{y+\alpha}+i A_n S_{\alpha}S_{y+\alpha}) {\hat {\bf
z}}\bigr\}&(3.7b)}$$
 where
$$\eqalignno{C_{\alpha} &=\cos\left({\alpha\pi\over 2}\right)\qquad C_{y+\alpha} =\cos\left(h y +{\alpha\pi\over 2}\right)\cr
S_{\alpha} &=\sin\left({\alpha\pi\over 2}\right)\qquad
S_{y+\alpha} =\sin\left(h y +{\alpha\pi\over 2}\right)}$$ $A_n,
C_n$ are the constant to be determined from initial conditions.
The fields given in equation~(3.7) have been plotted in Figure~2
for different values of $\alpha$ at an observation point
$(hy,\beta z)=(\pi/4,\pi/4)$.

 \vskip 4.0 in
 \hskip 0.5 in  \epsfxsize=15 pt\hskip -0.1 in\epsfbox[5 5 25 45] {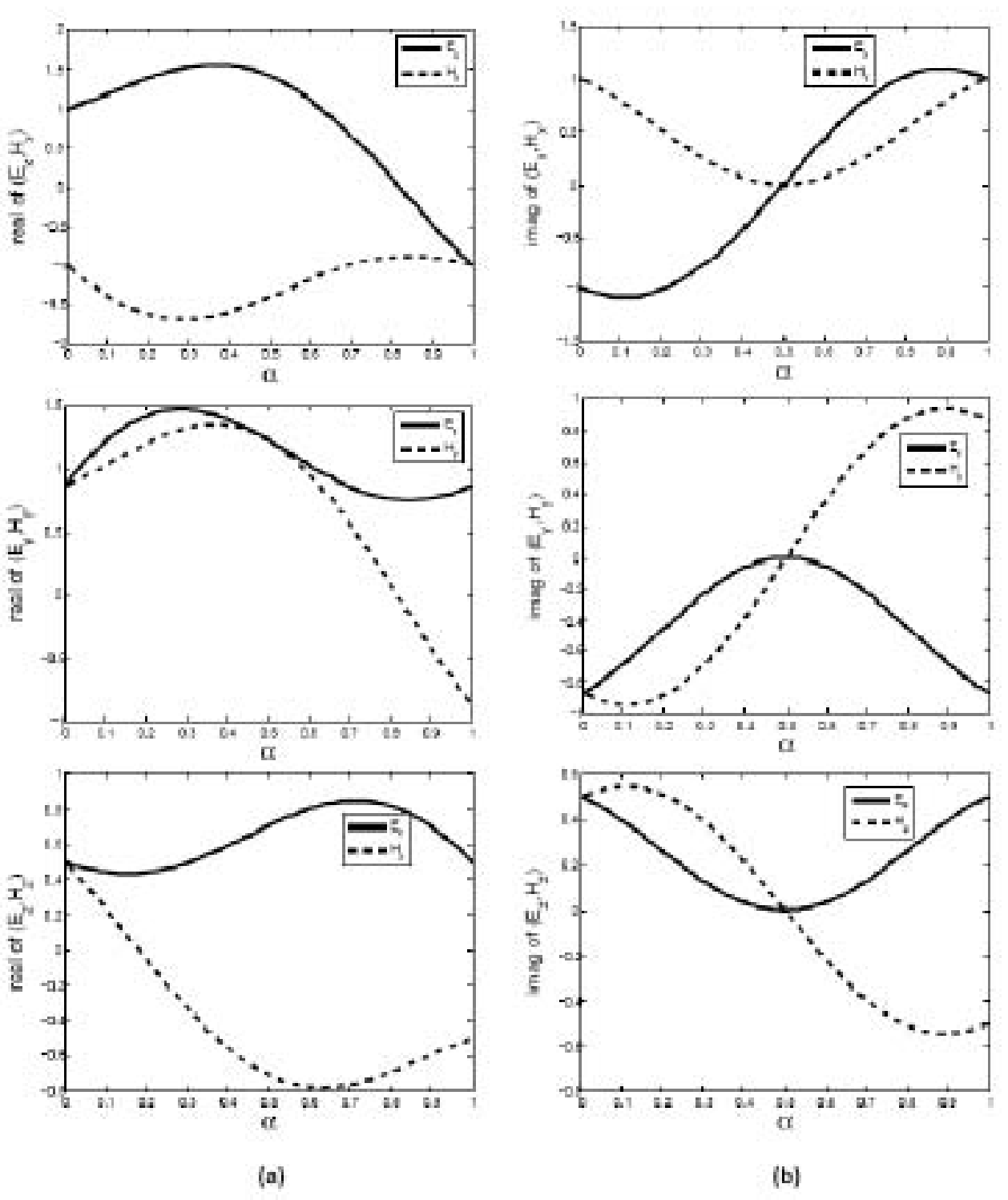}

 \hskip -0.6 in \caption Figure 2  ! \rm Plots of fractional dual
 $TE^z$ polarized fields at  a point  \rm $(h y, \beta  z)=(\pi/4,
\pi/4)$~\vskip -0.1 in ~~~$({\bf a})~\rm real~ parts ~~({\bf b})
~\rm imaginary~ parts$! \vfil\eject

From Figure~2, it can be seen that fractional dual fields satisfy
the principle of duality, i.e., for $\alpha=0$
$$\eqalignno{ {E}_{\rm fdx} &= {E}_{\rm x},\qquad \eta{H}_{\rm fdx} =\eta{H}_{\rm x} \cr
{E}_{\rm fdy} &= {E}_{\rm y},\qquad \eta{H}_{\rm fdy}
=\eta{H}_{\rm y} \cr {E}_{\rm fdz} &= {E}_{\rm z},\qquad
\eta{H}_{\rm fdz} =\eta{H}_{\rm z}}$$ and for $\alpha=1$
$$\eqalignno{{E}_{\rm fdx} &=
\eta{H}_{\rm x},\qquad \eta{H}_{\rm fdx} =-{E}_{\rm x} \cr
{E}_{\rm fdy} &= \eta{H}_{\rm y},\qquad \eta{H}_{\rm fdy}
=-{E}_{\rm y} \cr {E}_{\rm fdz} &= \eta{H}_{\rm z},\qquad
\eta{H}_{\rm fdz} =-{E}_{\rm z} }$$
\section{4. Results and discussion}
 \section{4.1 Behavior of fields inside the fractional parallel plate DB waveguide}
In order to study the behavior of fields inside the fractional
parallel plate DB waveguide, electric and magnetic field lines are
plotted in the yz-plane and are shown in Figure~3.

These plots are for the mode propagating through the guide at an
angle $\pi/6$ so that ${\beta / k}=\cos(\pi/6),~ {h /
k}=\sin(\pi/6)$. Initial conditions for both the modes are taken
as same. Solid lines show the electric as well as magnetic field
plots for DB waveguide while the fields of PEC waveguide are shown
by dashed lines as a reference. From the figure we see that there
is no normal component of the electric as well as magnetic field
for $\alpha=0$. This is because the plates of the guide behave as
perfect electric conductors for transverse electric components
while they behave as perfect magnetic conductor for transverse
magnetic modes. For the reference PEC results, there is no
tangential component of the electric filed at the guide surface
while magnetic field has no normal component. As value of the
$\alpha$ increase from 0, normal components of both the fields in
DB guide starts appearing and become maximum at $\alpha=0.5$.
After this value normal components start decreasing and again
become zero at $\alpha=1$. \vskip 5.0 in \hskip 0.5 in
 \epsfxsize=15 pt\hskip -0.2 in\epsfbox[5 5 25 45]
{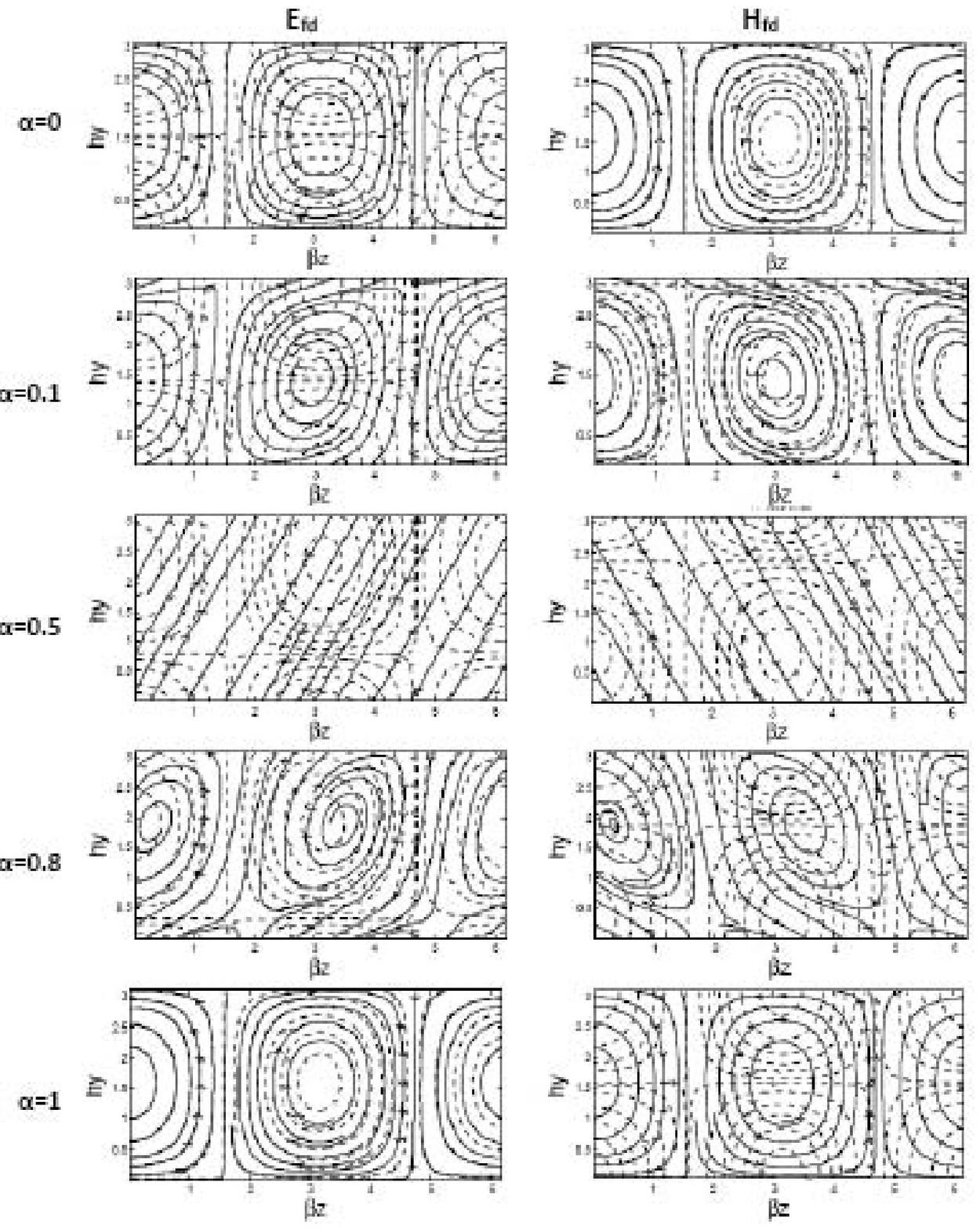} \vskip 0.0 in \caption Figure 3  ! \rm  Field lines in
yz-plane at different values of ${\alpha}$; solid lines are for the
fractional DB waveguides while dashed lines are for the fractional
PEC waveguides!

\vfil\eject
It may be noted that electric field distribution in
the DB waveguide is same as the magnetic field distribution for
the limiting values of $\alpha$ while it is different for the
intermediate values. Further it may be noted that field behavior
for the original and dual situation is similar.
\section{4.2 Transverse impedance of walls}

Wave impedance is defined by ratio of the transverse components of
the electric and magnetic fields as
$$\eqalignno{Z_{\rm fdxz}&=-{E_{\rm fdx }\over H_{\rm fdz }}=\eta{k \over
h}{{A_n S_{\alpha}C_{y+\alpha}+iC_n C_{\alpha}S_{y+\alpha}}\over
C_n C_{\alpha}C_{y+\alpha}+i A_n S_{\alpha}S_{y+\alpha}} \cr
Z_{\rm fdzx}&={E_{\rm fdz }\over H_{\rm fdx }}=\eta{h \over
k}{{A_n C_{\alpha}C_{y+\alpha}-i C_n S_{\alpha}S_{y+\alpha}}\over
{C_n S_{\alpha}C_{y+\alpha}-i A_n C_{\alpha}S_{y+\alpha}}}}$$ At
$y=0$, these impedances become impedance of the new reflecting
boundary called the fractional dual boundary. The normalized
impedance matrix of the DB boundary wall taking $A_n=C_n$ can be
written as
$$\eqalignno{\underline{\underline{z}}_{\rm fd}&=\left\{{k \over
h}z_{\rm fdxz}{\hat {\bf x}}{\hat {\bf z}} + {h \over k} z_{\rm
fdzx}{\hat {\bf z}}{\hat {\bf x}}\right\}, 0 \le \alpha \le 1}$$
where
$$\eqalignno{z_{\rm fdxz}&={{S_{\alpha}C_{\alpha}+i C_{\alpha}S_{\alpha}}\over
{C_{\alpha}C_{\alpha}+i S_{\alpha}S_{\alpha}}} \cr z_{\rm
fdzx}&={{C_{\alpha}C_{\alpha}-i S_{\alpha}S_{\alpha}}\over
{S_{\alpha}C_{\alpha}-i C_{\alpha}S_{\alpha}}}}$$ These impedance
components have been plotted for whole range of $\alpha$ as in
Figure~4. It may be noted that $z_{\rm fdxz}$ corresponds to the
impedance for $\rm TE^z$ modes and $z_{\rm fdzx}$ corresponds to
the impedance for $\rm TM^z$ modes. Since DB boundary behaves as
PEC for the $\rm TE^z$ modes so it is zero at $\alpha=0$ and
$\alpha=1$ while it is complex for the intermediate range of
$\alpha$. Similarly Since DB boundary behaves as PMC for the $\rm
TM^z$ modes so it is infinitely high at $\alpha=0$ and $\alpha=1$
while it is finite complex for the intermediate range of $\alpha$.
\vfil\eject $$\eqalignno{\cr}$$

 \vskip 0.8 in \hskip -0.5 in
 \epsfxsize=0 pt\hskip -0.0 in\epsfbox[5 5 25 45]
{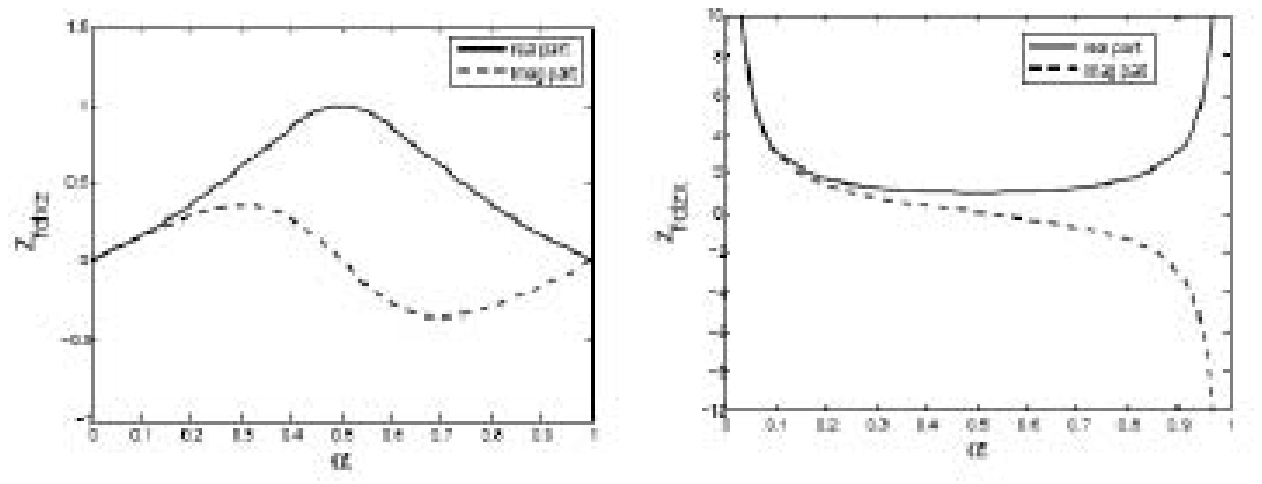} \vskip 0.0 in \caption Figure 4  ! \rm  Transverse
impedance of walls of fractional DB waveguides vs. fractional
parameter $\alpha$!
\section{5. Conclusions}
It may be concluded from the above results that fractional dual
solutions of the fields inside a parallel plate DB waveguide may
be derived using fractional curl operator. The waveguide modelled
by such fields may be termed as fractional dual parallel plate DB
waveguide. It has been noted that electric field distribution in
the DB waveguide is same as the magnetic field distribution for
the limiting values of $\alpha$ while it is different for the
intermediate values. Fractional dual waveguide may predict the
situation which is an intermediate step of DB boundary waveguide
and its dual situation. For limiting cases, transverse impedance
of DB wall is zero for transverse electric mode and infinitely
high for transverse magnetic mode while it is non zero complex
value for the intermediate situations.

\section{References}

\item{[1]} K. B. Oldham and J. Spanier, The Fractional Calculus,
Academic Press, New York, 1974.

\item{[2]} R. Hilfer, Applications of Fractional Calculus in
Physics, World Scientific, 2000.

\item{[3]} I. Podlubny, Fractional Differential Equations,
Mathematics in Science and Engineering V198, Academic Press, 1999.

\item{[4]} S. Das, Functional Fractional Calculus for System
Identification and Controls, Springer-Verlag, 2008.

\item{[5]} L. Debnath, Recent applications of fractional calculus
to science and engineering, International Journal of Mathematics
and Mathematical Sciences, Vol.~54, 3413-3442, 2003.

\item{[6]} N. Engheta, Note on fractional calculus and the image
method for dielectric spheres, Journal of Electromagnetic Waves
and Applications, Vol. 9, 1179-1188, 1995.

\item{[7]} N. Engheta, Use of fractional calculus to propose some
fractional solution for the scalar Helmholtzs equation, Progress
in Electromagnetics Research-PIER, Vol. 12, 107-132, 1996.

\item{[8]} N. Engheta, $\,$ Electrostatic  fractional image
methods for perfectly conducting wedges and cones,  IEEE
Transactions on Antennas and Propagation, Vol. 44, 1565-1574,
1996.

\item{[9]} N. Engheta, On the role of fractional calculus in
electromagnetic theory, IEEE Antennas and Propagation Magazine,
Vol. 39, 35-46, 1997.

\item{[10]} N. Engheta, Phase and amplitude of fractional-order
intermediate wave, Microwave and Optical Technology Letters, Vol.
21,  338-343, 1999..

\item{[11]} N. Engheta, Fractional Paradigm in Electromagnetic
Theory," a chapter in Frontiers in Electromagnetics-PIER, D. H.
Werner and R. Mittra (eds.), IEEE Press, chapter 12, pp. 523-552,
1999.

\item{[12]}V. E. Tarasov, Universal electromagnetic waves in
dielectric, J. Phys.: Condens. Matter, Vol. 20, 175-223, 2008.

\item{[13]}V. E. Tarasov, Fractional integro-differential
equations for electromagnetic waves in dielectric media, Teoret.
Mat. Fiz., Vol.~158, 419–-424, 2009.

\item{[14]} S. I. Musliha  and D. Baleanu, Fractional multipoles
in fractional space, Nonlinear Analysis: Real World Applications,
Vol. 8, 198-203, 2007.

\item{[15]} D. Baleanua,  A. K. Golmankhanehb,  and A. K.
Golmankhaneh, On electromagnetic field in fractional space,
Nonlinear Analysis: Real World Applications, Vol. 11, 288-292,
2010.

\item{[16]} M. Zubair, M. J. Mughal, Q. A. Naqvi, The wave
equation and general plane wave solutions in fractional space,
Progress In Electromagnetics Research Letters, PIERL, Vol. 19,
137-146, 2010.

\item{[17]} M. Zubair, M. J. Mughal, Q. A. Naqvi, A. A. Rizvi,
Differential electromagnetic equations in fractional space,
 Progress In Electromagnetics Research-PIER, Vol. 114, 255-269,
 2011.

\item{[18]} M. Zubair, M. J. Mughal, and Q. A. Naqvi, An exact
solution of the cylindrical wave equation for electromagnetic
field in fractional dimensional space, Progress In
Electromagnetics Research-PIER, Vol. 114, 443--455, 2011.

\item{[19]} M. Zubair,  M. J. Mughal, and Q.A. Naqvi, On
electromagnetic wave propagation in fractional space, Nonlinear
Analysis: Real World Applications, Vol. 12, 2844-2850, 2011.

\item{[20]} M. Zubair, M. J. Mughal, and Q. A. Naqvi, An exact
solution of the spherical wave equation in d-dimensional
fractional space, Journal of Electromagnetic Waves and
Applications, Vol. 25, 1481–-1491, 2011.

\item{[21]} N. Engheta, On Fractional paradigm and intermediate
zones in Electromagnetism: I. planar observation, Microwave and
Optical Technology Letters, Vol. 22, 236-241, 1999.

\item{[22]} N. Engheta, On Fractional paradigm and intermediate
zones in Electromagnetism: II. cylindrical and spherical
observations, Microwave and Optical Technology Letters, Vol. 23,
100-103, 1999.

\item{[23]} A. Lakhtakia, A representation theorem involving
fractional derivatives for linear homogeneous chiral media,
Microwave and Optical Technology Letters, Vol. 28, 385-386, 2001.

\item{[24]} N. Engheta, Fractional curl operator in
electromagnetics, Microwave and Optical Technology Letters, Vol.
17, 86-91, 1998.

\item{[25]} H. M. Ozaktas, Zee Zalevsky and M. A. Kutay, The
Fractional Fourier Transform with Applications in Optics and
Signal Processing, Wiley, New York, 2001.

\item{[26]} Q. A. Naqvi, and A. A. Rizvi, Fractional dual
solutions and corresponding sources, Progress in Electromagnetics
Research, PIER, 25,  223-238, 2000.

\item{[27]} Q. A. Naqvi, G. Murtaza, and A. A. Rizvi, Fractional
dual solutions to Maxwell equations in homogeneous chiral medium,
Optics Communications, Vol. 178, 27-30, 2000.

\item{[28]} Q. A. Naqvi, and M. Abbas, Complex and higher order
fractional curl operator in electromagnetics, Optics
Communications, 241, 349-355, 2004.

\item{[29]} Q. A. Naqvi, and M. Abbas, Fractional duality and
metamaterials with negative permittivity and permeability, Optics
Communications, 227, 143-146, 2003.

\item{[30]} E. I. Veliev, M. V. Ivakhnychenko, and T. M. Ahmedov,
Fractional boundary conditions in plane waves diffraction on a
strip, Progress In Electromagnetics Research-PIER, 79, 443--462,
2008.

\item{$[31]$} {A. Hussain and Q. A. Naqvi,}
              {Fractional curl operator in
              chiral medium and fractional nonsymmetric transmission line, Progress
In Electromagnetics Research-PIER, 59, 199-213, 2006.}

\item{$[32]$} {A. Hussain, S. Ishfaq and Q. A. Naqvi,}
              {Fractional curl operator and fractional waveguides, Progress
In Electromagnetics Research-PIER, 63, 319-335, 2006.}

\item{$[33]$} {A. Hussain, M. Faryad and Q. A. Naqvi,} Fractional
curl operator and fractional chiro-waveguide, Journal of
Electromagnetic Waves and Applications, Vol.~21, 1119-1129, 2007.

\item{[34]} M. Faryad and Q. A. Naqvi, Fractional rectangular
waveguide, Progress In Electromagnetics Research-PIER, Vol.~75,
383-396, 2007.

\item{[35]} A. Hussain and Q. A. Naqvi, Perfect electromagnetic
conductor (PEMC) and fractional waveguide, Progress In
Electromagnetics Research-PIER, Vol. 73, 61-69, 2007.

\item{[36]} H. Maab and Q. A. Naqvi, Fractional surface waveguide,
Progress In Electromagnetics Research C, PIERC, Vol. 1, 199-209,
2008.

\item{[37]} A. Hussain and Q. A. Naqvi, $\,$ Fractional
rectangular impedance waveguide, Progress In Electromagnetics
Research-PIER, Vol. 96, 101-116, 2009.

\item{[38]} H. Maab and Q. A. Naqvi, Fractional rectangular cavity
resonator, Progress In Electromagnetics Research B, PIERB, Vol. 9,
69-82, 2008.

\item{[39]} {A. Hussain, M. Faryad and Q. A. Naqvi,}
              {Fractional waveguides with impedance walls, Progress in Electromagnetic Research C, PIERC, Vol.~4, 191-204, 2008.}

\item{[40]} A. Hussain and Q. A. Naqvi,  Fractional rectangular
impedance $\,$ waveguide, Progress in Electromagnetics
Research-PIER, Vol.~96, 101-116, 2009.

\item{[41]} {S. A. Naqvi,  Q. A. Naqvi,  and A. Hussain,}
{Modelling of transmission through a chiral slab using fractional
curl operator, Optics Communications, 266, pp:~404-406, 2006.}

\item{[42]} S. A. Naqvi, M. Faryad, Q. A. Naqvi, and M. Abbas,
Fractional duality in homogeneous bi-isotropic medium, Progress In
Electromagnetics Research-PIER, 78, 159–-172, 2008

\item{[43]} A. Lakhtakia, An electromagnetic trinity from
``Negative permittivity" and "Negative permeability". Int. Journal
of Infrared and Millimeter Waves, Vol. 22, 1731- 1734, 2001.

\item{[44]} S. Tretyakov, I. Nefedov, A. Sihvola, S. Maslovski, A
Metamaterial with Extreme Properties: The Chiral Nihility,
Progress in Electromagnetics Research Symposium 2003, p. 468,
Honolulu, Hawaii, USA, October 13-16, 2003.

\item{[45]} S. Tretyakov, I. Nefedov, A. Sihvola, S. Maslovski, C.
Simovski, Waves and energy in chiral nihility, Journal of
Electromagnetic Waves and Applications, Vol.~17, 695-706, 2003.

\item{[46]} S. A. Tretyakov, I. S. Nefedov, P. Alitalo,
Generalized field transforming metamaterials, New Journal of
Physics, 10, 115028, 2008.

\item{[47]} Q. Cheng,  T. J. Cui,  C. Zhang, Waves in planar
waveguide containing chiral nihility metamaterial, Optics
Communications, 276 317-–321, 2007.

\item{[48]} C. Zhang and T. J. Cui, Negative reflections of
electromagnetic waves in chiral media, arXiv:physics/0610172.

\item{[49]} J. F. Dong and C. Xu, Surface polaritons in planar
chiral nihility meta-material waveguides, Optics Communications,
Vol. 282, 3899-3904, 2009.

\item{[50]} A. Naqvi, Comments on waves in planar waveguide
containing chiral nihility metamaterial, Optics Communications,
Vol. 284, 215-216, 2011.

\item{[51]} Q. A. Naqvi, Fractional dual solutions to the Maxwell
equations in chiral nihility medium, Optics Communications, Vol.
282, 2016-2018, 2009.

\item{[52]} Q. A. Naqvi, Planar slab of chiral nihility
metamaterial backed by fractional dual/PEMC interface, Progress In
Electromagnetics Research-PIER, Vol. 85, 381–391, 2009.

\item{[53]}  Q. A. Naqvi, Fractional dual solutions in grounded
chiral nihility slab and their effect on outside fields, J. of
Electromagnetic Waves and Applications, JEMWA, Vol. 23, 5/6,
773--784, 2009

\item{[54]} A. Naqvi, S. Ahmed, and Q. A. Naqvi, Perfect
electromagnetic conductor and fractional dual interface placed in
a chiral nihility medium, Journal of Electromagnetic Waves and
Applications, Vol. 24, 1991-1999, 2010.

\item{[55]}  A. Illahi and Q. A. Naqvi, Study of focusing of
electromagnetic waves reflected by a PEMC backed chiral nihility
reflector using Maslov's method, Journal of Electromagnetic Waves
and Applications, JEMWA,  Vol. 23,  863--873, 2009

\item{[56]} Q. A. Naqvi, Fractional Dual Interface in Chiral
Nihility Medium, Progress In Electromagnetics Research Letters,
PIERL, Vol. 8, 135-142, 2009.

\item{[57]} A. Naqvi, A. Hussain, and Q. A. Naqvi,  Waves in
fractional dual planar waveguides containing chiral nihility
metamaterial, Journal of Electromagnetic Waves and Applications,
Vol. 24, 1575-1586, 2010.

\item{[58]} I. V. Lindell, A. H. Sihvola, Zero axial parameter
(ZAP) sheet, Progress In Electromagnetics Research-PIER, Vol. 89,
213-224, 2009.

\item{[59]} I. V. Lindell, A. H. Sihvola, Uniaxial IB-medium
interface and novel boundary conditions, IEEE Trans. Antennas
Propagat., Vol. 57, 694-700, 2009.

\item{[60]} I. V. Lindell and A. Sihvola,  Circular waveguide with
DB boundary Conditions, IEEE Trans. on Micro. Theory and Tech.,
Vol. 58, 903-909, 2010.

\item{[61]} I. V. Lindell,  H. Wallen, and  A. Sihvola,   General
 electromagnetic boundary conditions involving normal field
 components, IEEE Ant. and Wirel. Propag. Lett., Vol. 8, 877-880, 2009.

\item{[62]} A. H. Sihvola, Wallen, and P. Yla-Oijala, Scattering
by DB spheres, IEEE Ant. and Wirel. Propag. Lett., Vol. 8,
542-545, 2009.

\item{[63]} I. V. Lindell and A. Sihvola, Electromagnetic boundary
and its realization with anisotropic metamaterial, Phys. Rev. E,
Vol. 79, 026604, 2009.

\item{[64]} I. V. Lindell and A. Sihvola,  Zero - axial-parameter
 (Zap) medium sheet,  Progress In Electromagnetics Research-PIER,
    Vol. 89, 213-224, 2009.

\item{[65]} A. Naqvi, F. Majeed, and Q. A. Naqvi, Planar db
boundary placed in a chiral and chiral nihility metamaterial,
Progress In Electromagnetics Research Letters, PIERL, Vol. 21,
41-48, 2011.

 \bye